\documentstyle[sprocl]{article}

\begin{document}

\begin{flushright}
IUHET 335\\
May 1996
\end{flushright}
\medskip
\title{CPT, STRINGS, AND NEUTRAL MESONS}
\author{V.A.~KOSTELECK\'Y}
\address{Physics Department, 
Indiana University, Bloomington, IN 47405, U.S.A.}


\maketitle
\abstracts{
A sketch of our results on CPT and strings is given.
A mechanism for spontaneous CPT violation in string theory
is briefly reviewed,
and recent theoretical progress is summarized.
Possible CPT-violating contributions to a four-dimensional
low-energy effective theory could produce
effects at levels testable in neutral-meson systems.
Likely experimental signatures and attainable bounds
are outlined in candidate experiments with correlated and
uncorrelated neutral mesons.}

\medskip
\def\al{\alpha}
\def\be{\beta}
\def\ga{\gamma}
\def\de{\delta}
\def\ep{\epsilon}
\def\ve{\varepsilon}
\def\ze{\zeta}
\def\et{\eta}
\def\th{\theta}
\def\vt{\vartheta}
\def\io{\iota}
\def\ka{\kappa}
\def\la{\lambda}
\def\vpi{\varpi}
\def\rh{\rho}
\def\vr{\varrho}
\def\si{\sigma}
\def\vs{\varsigma}
\def\ta{\tau}
\def\up{\upsilon}
\def\ph{\phi}
\def\vp{\varphi}
\def\ch{\chi}
\def\ps{\psi}
\def\om{\omega}
\def\Ga{\Gamma}
\def\De{\Delta}
\def\Th{\Theta}
\def\La{\Lambda}
\def\Si{\Sigma}
\def\Up{\Upsilon}
\def\Ph{\Phi}
\def\Ps{\Psi}
\def\Om{\Omega}
\def\mn{{\mu\nu}}
\def\cl{{\cal L}}
\def\fr#1#2{{{#1} \over {#2}}}
\def\prt{\partial}
\def\ap{\al^\prime}
\def\apt{\al^{\prime 2}}
\def\apth{\al^{\prime 3}}
\def\pt#1{\phantom{#1}}
\def\vev#1{\langle {#1}\rangle}
\def\bra#1{\langle{#1}|}
\def\ket#1{|{#1}\rangle}
\def\bracket#1#2{\langle{#1}|{#2}\rangle}
\def\expect#1{\langle{#1}\rangle}
\def\sbra#1#2{\,{}_{{}_{#1}}\langle{#2}|}
\def\sket#1#2{|{#1}\rangle_{{}_{#2}}\,}
\def\sbracket#1#2#3#4{\,{}_{{}_{#1}}
 \langle{#2}|{#3}\rangle_{{}_{#4}}\,}
\def\sexpect#1#2#3{\,{}_{{}_{#1}}\langle{#2}\rangle_{{}_{#3}}\,}
\def\half{{\textstyle{1\over 2}}}
\def\frac#1#2{{\textstyle{{#1}\over {#2}}}}
\def\ni{\noindent}
\def\lsim{\mathrel{\rlap{\lower4pt\hbox{\hskip1pt$\sim$}}
    \raise1pt\hbox{$<$}}}
\def\gsim{\mathrel{\rlap{\lower4pt\hbox{\hskip1pt$\sim$}}
    \raise1pt\hbox{$>$}}}
\def\sqr#1#2{{\vcenter{\vbox{\hrule height.#2pt
         \hbox{\vrule width.#2pt height#1pt \kern#1pt
         \vrule width.#2pt}
         \hrule height.#2pt}}}}
\def\square{\mathchoice\sqr66\sqr66\sqr{2.1}3\sqr{1.5}3}

\newcommand{\beq}{\begin{equation}}
\newcommand{\eeq}{\end{equation}}
\newcommand{\bea}{\begin{eqnarray}}
\newcommand{\eea}{\end{eqnarray}}
\newcommand{\rf}[1]{(\ref{#1})}

\baselineskip=11pt
CPT invariance 
is a fundamental theoretical property of
local relativistic field theories of point particles.\cite{cpt}
Moreover, 
interferometric experiments with neutral kaons
have tested it to a high degree of precision.\cite{pdg}
These features make CPT violation
an ideal candidate signature for new,
non-particle physics.\cite{kp1}

Currently,
string theory is among the most promising frameworks
for a unified description of the fundamental forces.
Since strings are extended objects,
they do not automatically satisfy the axioms
leading to CPT invariance.
Indeed,
an explicit stringy mechanism for CPT breaking is known,\cite{kp1}
connected to spontaneous Lorentz breaking.\cite{ks}
Briefly,
it turns out that couplings in string field theory exist 
that are impossible in conventional four-dimensional 
renormalizable gauge theories.
When scalars acquire expectation values,
these couplings can induce instabilities in
effective potentials for Lorentz tensors.
Nonzero expectation values for the latter can ensue,
leading under suitable circumstances to CPT violation.

Additional support for this scenario is provided by a recent 
detailed analysis of Lorentz- and CPT-breaking solutions 
of the equations of motion for the open bosonic string.\cite{kp2}
Dominant terms of the action in a level-truncation scheme 
are computed analytically,
and the equations of motion are solved.
For some solutions,
over 20,000 nonvanishing terms in the action are used.
The associated nonzero expectation values 
for various tensor fields are found to exhibit
predicted features.

If a string theory indeed describes the Universe,
then the above mechanism might induce CPT-violating terms
in the effective four-dimensional low-energy theory,
i.e., in the standard model.
Since no zeroth-order CPT violation is observed,
it must be suppressed.
The natural suppression factor is the ratio 
$\sim 10^{-17}$ 
of the low-energy scale to the Planck scale.
It turns out that a suppression of this magnitude
can produce effects around or just below those
detectable in current experiments 
with neutral-meson systems.\cite{kp1,kps}

The issues of modeling and detecting these effects 
have been examined\cite{kps,ck,kk}
in the $K$ system, the $D$ system, and the two $B$ systems.
Candidate experiments include ones 
with uncorrelated mesons or with correlated mesons
(factories).
Several interesting features emerge from the analyses.

$\bullet$
In the string scenario,
the magnitude of possible CPT violation depends 
on coupling constants that can vary among different quark flavors
as, for example, do the usual Yukawa couplings.
This means it would be desirable to test for CPT violation 
in all the neutral-meson systems, 
not only the $K$ system.

$\bullet$
The only bounds in the literature to date are those
on the components of the
$K$-system CPT-violating parameter $\de_K$.
These may be improved soon,
e.g., in $\ph$ factories,
under favorable circumstances
possibly to one part in $10^{5}$.

$\bullet$
Testing CPT in the $D$ system is hard because the
mixing is so small.
Nonetheless,
definite tests can be performed at a $\ta$-charm factory.
They could include bounding the $K$-system parameter $\de_K$ 
and the corresponding quantity $\de_D$ in the $D$ system.

$\bullet$
Testing CPT in the neutral-$B_d$ system is particularly
interesting because it involves the $b$ quark,
and in some string scenarios CPT violation is largest 
in this system.
Although no bound is currently in the literature,
sufficient data already exist 
to bound the CPT-violating parameter $\de_B$ at about
the 10\% level.
Prospects at the $B$ factories appear interesting.

\medskip

My thanks to Don Colladay, Rob Potting, Stuart Samuel, 
and Rick Van Kooten for enjoyable collaborations leading 
to the results described here.



\medskip

\begin{thebibliography}{xx}

\bibitem{cpt}
See, e.g., R.F. Streater and A.S. Wightman,
\it PCT, Spin and Statistics, and All That, \rm
Benjamin Cummings, Reading, 1964.

\bibitem{pdg}
See, e.g., Review of Particle Properties,
Phys.~Rev.~D {\bf 50}, 1173 (1994).

\bibitem{kp1}
V.A. Kosteleck\'y and R. Potting,
Nucl.~Phys.~B {\bf 359} (1991) 545.

\bibitem{ks}
V.A. Kosteleck\'y and S. Samuel,
Phys.~Rev.~D {\bf 39} (1989) 683;
\it ibid., \rm
{\bf 40} (1989) 1886;
Phys.~Rev.~Lett. {\bf 63} (1989) 224;
\it ibid., \rm
{\bf 66} (1991) 1811.

\bibitem{kp2}
V.A. Kosteleck\'y and R. Potting,
Phys.~Lett.~B, in press (hep-th/9605088).

\bibitem{kps}
V.A.~Kosteleck\'y and R.~Potting,
Phys.~Rev.~D {\bf 51}, 3923 (1995);
see also V.A.~Kosteleck\'y, R.~Potting and S.~Samuel,
in S.~Hegarty et al., eds.,
\it Proc.~1991~Joint~Intl.~Lepton-Photon~Symp.~ 
and~Europhys.~ Conf.~ on High Energy Phys.~\rm
(World Scientific 1992);
V.A.~Kosteleck\'y and R.~Potting,
in D.B.~Cline, ed.,
\it Gamma Ray--Neutrino Cosmology and Planck Scale Physics \rm
(World Scientific 1993)
(hep-th/9211116).

\bibitem{ck}
D. Colladay and V.A. Kosteleck\'y,
Phys.~Lett.~B {\bf 344}, 259 (1995);
Phys.~Rev.~D {\bf 52}, 6224 (1995).

\bibitem{kk}
V.A. Kosteleck\'y and R. Van Kooten,
Phys.~Rev.~D, in press (hep-ph/9607449).

\end{thebibliography}
\end{document}